\documentclass{article}

\usepackage{PRIMEarxiv}

\usepackage[utf8]{inputenc} 
\usepackage[T1]{fontenc}    
\usepackage{hyperref}       
\usepackage{url}            
\usepackage{booktabs}       
\usepackage{amsfonts}       
\usepackage{nicefrac}       
\usepackage{microtype}      
\usepackage{lipsum}
\usepackage{fancyhdr}       
\usepackage{graphicx}       
\graphicspath{{media/}}     

\usepackage{amssymb}
\usepackage{amsmath}
\usepackage[table,xcdraw]{xcolor}
\usepackage{multirow}

\title{Improving Portfolio Performance Using a Novel Method for Predicting Financial Regimes
}

\author{
  Piotr Pomorski, Denise Gorse \\
  Department of Computer Science \\
  University College London \\
  London\\
  \texttt{\{P.Pomorski, D.Gorse\}@cs.ucl.ac.uk} \\
}

\begin{document}
\maketitle

\begin{abstract}
This work extends a previous work in regime detection, which allowed trading positions to be profitably adjusted when a new regime was detected, to ex ante prediction of regimes, leading to substantial performance improvements over the earlier model, over all three asset classes considered (equities, commodities, and foreign exchange), over a test period of four years. The proposed new model is also benchmarked over this same period against a hidden Markov model, the most popular current model for financial regime prediction, and against an appropriate index benchmark for each asset class, in the case of the commodities model having a test period cost-adjusted cumulative return over four times higher than that expected from the index. Notably, the proposed model makes use of a contrarian trading strategy, not uncommon in the financial industry but relatively unexplored in machine learning models. The model also makes use of frequent short positions, something not always desirable to investors due to issues of both financial risk and ethics; however, it is discussed how further work could remove this reliance on shorting and allow the construction of a long-only version of the model.
\end{abstract}


\section{Introduction}
Systematic methodologies that attempt to exploit patterns in financial time series have been around for a long time, in the US dating back to the late 1800s and the work of Charles Dow (the creator of Dow Jones Industrial Average), and in Japan and the Netherlands even longer. However, this field, referred to in the financial industry as \textit{technical analysis} (TA) fell into disrepute in the 1980s due to the rise of the Efficient Markets Hypothesis (EMH),
which claims it is impossible to `beat the market' using past data, since the stock price already incorporates historical and current information. 
But the EMH has since itself been challenged by the emerging field of behavioural finance, and the work of Andrew Lo and others~\cite{lo2011non} has evidenced that there are indeed exploitable signals buried in the noise presented by financial time series data.
A further reason for this recovered optimism has been the rise of machine learning (ML) models for financial time series forecasting, able to discover weaker, and subtler, signals than could otherwise be detected. While some ML models use deep learning to uncover patterns in raw data, the majority use engineered features, including ones derived from TA, to reduce the dimensionality of the data and add value to it, with the strongest-evidenced (for example, in~\cite{pavlov2012testing}) TA tools being of the moving average type. Moving averages do, however, have issues with the choice of window length, though these can be overcome by the use of adaptive moving averages, in particular \textit{Kaufman's adaptive moving average} (KAMA)~\cite{kaufman1995smarter}.

Recently, in~\cite{pomorski2023improving}, KAMA, adept in trend (upward, or `bullish', versus downward, or `bearish') detection, was combined with the  \textit{two-state Markov switching regression} (MSR) model~\cite{krolzig1997markov}, as a means of distinguishing volatility regimes, in order to separate financial time series into four states: low-variance (LV) bullish, low-variance (LV) bearish, high-variance (HV) bullish, and high-variance (HV) bearish, with the detected regimes then used by the resulting \textit{KAMA+MSR} model as the basis of a profitable trading strategy.
However, any strategy based on regime detection will be subject to inevitable losses, due to having remained in a now-disadvantageous trading position after the market has switched regime, or due to not having sufficiently quickly taken advantage of a new trading opportunity.
This work therefore builds on the work of~\cite{pomorski2023improving}, by using KAMA+MSR to appropriately label periods of the training data, these labels being then used as targets for a random forest (RF) predictor, resulting in what we term the \textit{KMRF} (KAMA+MSR+RF) model. It will be demonstrated that the KMRF regime predictor does indeed improve substantially on results obtainable using KAMA+MSR (utilising cost-adjusted returns, with realistically-estimated transaction costs), with respect to a range of financial metrics.

\section{Machine learning for financial regime prediction}\label{sec:related_work}

Regime prediction, like most other areas of financial prediction, is nowadays dominated by machine learning (ML) models, traditional tools such as the Probit model~\cite{aldrich1984linear}, used in the 1990s to predict economic recessions in the US and elsewhere, having now been set aside, as even the simpler ML methods, such as elastic net, have been shown more effective~\cite{vrontos2021modeling}. 
The dominant ML model for regime prediction in industry and academia is the hidden Markov model (HMM)~\cite{baum1966statistical}, used for example to infer future price trends (downward and upward) in the oil market~\cite{e2010forecasting}, Euro/USD and AUD/USD crosses~\cite{lee2014hidden}, as well as Bitcoin~\cite{giudici2020hidden}. The HMM however suffers from an inability to predict regime switches without experiencing the onset of the new regime: the HMM can in effect only predict \textit{continuations} of already-changed regimes. Due to this limitation, many other ML models have been explored for regime prediction (though deep learning models have been less-used due to a relative scarcity of data and consequent risk of overfitting these complex models), with random forest (RF)~\cite{breiman2001random} standing out as the most promising model for regime prediction.

RF models have been used successfully in many areas of finance.
For example, RF was shown in~\cite{ballings2015evaluating} 
to be the most effective algorithm for prediction of one-year ahead price direction, for over 5500 European stocks, on the basis of fundamental and macro variables, compared with those made by a number of other algorithms, such as SVM, ANN, and k-nearest neighbours. Furthermore, when compared to SVM and na\"{i}ve Bayes in~\cite{milosevic2016equity},
RF achieved the highest F-Score in stock price correction (at least 10\% upward or downward movement) prediction over the long term. 
Moving to regime prediction, RF models have been used, for example, to predict bank crises~\cite{ward2017spotting}, regime turning points~\cite{piger2020turning} (in which RF outperformed gradient-boosted trees and na\"{i}ve Bayes), US recessions~\cite{yazdani2020machine} (again outperforming gradient-boosted trees, as well as support vector machine and ANN models), and in the prediction of stock market regimes~\cite{uysal2021machine}. 

It was thus decided to use the RF model to predict the assigned KAMA+MSR labels, with the HMM model, due to its long history of use for regime prediction, as a predictor benchmark (alongside the KAMA+MSR model~\cite{pomorski2023improving}, in order to establish the advantage of regime prediction over detection). 

\section{Methodology}\label{sec:methodology}

\begin{figure}
\centering
\includegraphics[scale = 0.45]{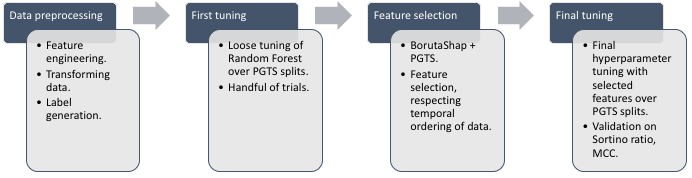}
\caption{Summary of the methodology adopted in this work.} \label{fig:methodology_flowchart}
\end{figure}

A summary flowchart of our methodology is given at the conclusion of this section in Figure~\ref{fig:methodology_flowchart}, with the individual components within this workflow discussed in more detail in the subsections below.

\subsection{Data and feature engineering}\label{sec:data}

\subsubsection{Data used}

Regimes will be predicted for three asset classes: equities, commodities, and foreign exchange (FX) pairs. The component assets for each class are listed online~\cite{features_and_assets}, as well as the start dates of the data, after first having gone through scaling process known as 
\textit{fractional differencing}~\cite{hosking1981fractional}, which balances stationarity and non-stationarity within time series.
(It should be noted that it is not possible to release the datasets per se, for commercial reasons; the code used also cannot be shared for these reasons.) 
Note that each asset class has a separate benchmark used only in out-of-sample model performance comparison, and that all series, including the benchmarks, end on the same date, 29/04/2022.

\subsubsection{Candidate features}\label{sec:candidate_features}

A wide range of features, also listed online~\cite{features_and_assets}, were considered as a pool from which the feature selection method of Section~\ref{sec:feature_selection} could draw. 
These features fall into three broad categories: \textit{technical} (statistical measures created from price series), \textit{fundamental} (measures related to a company's financial health), and \textit{macroeconomic} (related to wider economic forces), with the
technical features, 60 in total, used in the models for all three asset classes, computed using the TA~\footnote{\url{https://github.com/bukosabino/ta}. Last accessed 29 March 2023.} (28 features) and tsfresh~\cite{christ2016distributed} (32 features) packages.

\subsection{Transaction costs} 

Both hyperparameter optimisation and out-of-sample performance assessment will use measures based on cost-adjusted returns (Eq.~\ref{eq:cost_adjusted_returns}). 
The costs assumed in this work 
are listed in Table~\ref{tab:transaction_costs}; these are realistic estimates which were adopted after appropriate consultation with industry experts. 

\begin{table}[] 
\caption {Two-way (buy (long), sell (short)) trading costs for each asset class}
\centering
\begin{tabular}{|l|c|c|c|c|}
\hline
\multicolumn{1}{|c|}{\textbf{Asset class}} & \textbf{\begin{tabular}[c]{@{}c@{}}Brokerage  \\ commissions (\%)\end{tabular}} & \textbf{\begin{tabular}[c]{@{}c@{}}Bid-ask\\ spread (\%)\end{tabular}} & \textbf{\begin{tabular}[c]{@{}c@{}}Market\\ impact (\%)\end{tabular}} & \textbf{\begin{tabular}[c]{@{}c@{}}Total \\ cost (\%)\end{tabular}} \\ \hline
Equities                                   & 0.07                                                                            & 0.065                                                                   & 0.265                                                                  & 0.40                                                                 \\ \hline
Commodities                                & 0.14                                                                            & 0.13                                                                   & 0                                                                     & 0.27                                                                \\ \hline
Currencies (FX)                                & 0                                                                               & 0.13                                                                   & 0                                                                     & 0.13                                                                \\ \hline
\end{tabular}
\label{tab:transaction_costs}
\end{table}

\subsection{Hyperparameter optimisation} \label{sec:hyperparameter_optimisation} 

Hyperparameter optimisation (using the Optuna~\cite{akiba2019optuna} package) for the RF component of the three KMRF models was done in two phases, with the objective of optimising a financially-relevant metric, the Sortino ratio of Section~\ref{sec:financial_performance_metrics}. 
Search ranges and optimal hyperparameter values are listed in Table~\ref{tab:hyperparameters}, other RF hyperparameters, deemed of lesser importance, being set to their default values in scikit-learn~\footnote{\url{https://scikit-learn.org/stable/}. Last accessed 6 September 2022.}. 
The first phase consisted of a loose optimisation, in order to get an adequately working model for each asset class; feature selection was then done, as will be described below, followed by a second phase of more rigorous and computationally demanding hyperparameter optimisation. 85\% of each time series was used for training / validation (except in the case of the benchmarks) and 15\% was used for testing. 
In the case of time series, however, cross-validation needs to respect temporal ordering; in addition, care should be taken to avoid data leakage that might arise due to the computation of lagged statistical features. 

The \textit{purged group time-series split}  (PGTS) method~\footnote{\url{https://www.kaggle.com/code/marketneutral/purged-time-series-cv-xgboost-optuna}. Last accessed 12 July 2023.}, recommended in~\cite{de2018advances}, is used here.
PGTS is a cross-validation technique for time series that avoids data leakage. It uses non-overlapping groups to allow multiple time series as inputs, though these may start on different dates. During cross-validation, PGTS generates multiple splits, with groups remaining distinct across folds, with the training sets progressively including more data, maintaining temporal relationships. 

\begin{table}[] 
\caption {Hyperparameter search ranges and optimised values for each of the three RF models (equities, commodities, and currencies (FX)) built.}
\centering
\begin{tabular}{|l|c|c|c|c|}
\hline
\multicolumn{1}{|l|}{\textbf{Hyperparameter}} & \textbf{\begin{tabular}[c]{@{}c@{}}Search space\end{tabular}} & \textbf{\begin{tabular}[c]{@{}c@{}}Equities\end{tabular}} & \textbf{\begin{tabular}[c]{@{}c@{}}Commodities\end{tabular}} & \textbf{\begin{tabular}[c]{@{}c@{}}Currencies\end{tabular}} \\ \hline
\textit{n\_estimators}                                   & \{10,...,300\}                                                                            & 220 & 280 & 240  \\ \hline
\textit{max\_depth}                                & \{1,...,20\}                                                                            & 13 & 3 & 7  \\ \hline
\textit{min\_samples\_split}                                & \{1,...,100\}                                                                              & 76 & 18 & 22   \\ \hline
\textit{min\_samples\_leaf}                                & \{1,...,100\}                                                                              & 95 & 95 & 60   \\ \hline
\textit{max\_samples}                                & [0.1,...,1.0]                                                                              & 0.3649 & 0.1247 & 0.3603   \\ \hline
\textit{min\_weight\_fraction\_leaf}                                & [0.0,...,0.05]                                                                              & 0.0454 & 0.0213 & 0.0349   \\ \hline
\textit{max\_features}                                & [0.2,...,1.0]                                                                              & 0.2481 & 0.4001 & 0.3410   \\ \hline
\end{tabular}
\label{tab:hyperparameters}
\end{table}

\subsection{Feature selection} \label{sec:feature_selection} 

While RF has its own, built-in, feature selection tool, in the form of \textit{mean decrease in impurity} (MDI), it has been argued that this is not ideal as a means to identify the most important features (see, for example, \cite{strobl2007bias} and \cite{kursa2010feature}). 
Feature importance is therefore investigated here using the \textit{BorutaShap}~\footnote{\url{https://github.com/Ekeany/Boruta-Shap}. Last accessed 5 April 2023.} package. 
BorutaShap did, however, need adaptation for use with time series data, and the original code was modified to use PGTS cross-validation, described previously, in order to avoid data leakage; note that the same validation splits were used in loose, initial tuning, and in final optimisation. 

\subsection{Regime label generation} 

As noted in the Introduction, KAMA+MSR~\cite{pomorski2023improving} creates four regime classes: low-variance (LV) bullish, LV bearish, high-variance (HV) bullish, and HV bearish, though with only the first and the fourth of these being used for trading.  
The three (one-hot encoded) target labels used by the proposed KMRF model are derived from these four KAMA+MSR classes as below:

\begin{itemize}
    \item \textit{Bullish} = LV bullish + extension to the peak of next HV bullish regime.
    \item \textit{Bearish} = HV bearish + extension to the trough of next LV bearish regime.
    \item \textit{`Other'} = remaining parts of the HV bullish and LV bearish regimes.
\end{itemize}

\noindent The above extensions are carried out because HV bullish and LV bearish states contain periods of up-trending and down-trending markets, respectively. For example, after an LV bullish period the price of a certain asset, such as a commodity, may soar significantly and then suddenly drop, thus effectively entering an HV bullish state due to the volatile movement of the market. Similarly, subsequent to an HV bearish regime, an LV bearish state may occur when the volatility of the asset’s price becomes milder, and it initially falls but afterward rebounds (a typical `bear rally'). Such periods, if classified as `other', could confuse the learning algorithm, as well as potentially result in lesser profits.
Finally, any period during which a bullish or bearish label, even if correctly predicted, could not be usefully exploited because the transaction costs of Table \ref{tab:transaction_costs} would outweigh the profits made, is converted to the `other' class.

\subsection{Contrarian interpretation of predicted regime labels} \label{sec:contrarian_interpretation}

Initially, it was assumed that the models’ predictions would be used to establish a trend-following strategy, that is, a high probability of a bullish (prices rising) regime would give a buying signal, while a high probability of a bearish (prices falling) regime would give a shorting signal. (A high probability of the `other' regime after previously having entered a long or short position in an asset would indicate a point of time to close this position.)
However, preliminary results indicated that the predictions for the KMRF models in fact produced contrarian signals. In other words, a high probability of a bullish regime here indicated an exhausted and overbought market, which should spur an action to enter a short position in the underlying asset, while a high probability of a bearish regime pointed at oversold market conditions, thus recommending entering a long position in the underlying asset. Contrarian models are commonly used in the industry; many trading strategies act on, and many indicators provide an alarm about, overbought and oversold conditions of the market, and recommend assuming a contrarian position, shorting when overbought or buying when oversold. Following the discovery that a contrarian interpretation of the models' signals was advisable, the following strategy was therefore adopted:

\begin{itemize}
    \item When a bullish regime is predicted, a short position should be assumed in the underlying asset.
    \item When a bearish regime is predicted, a long position should be assumed in the underlying asset.
    \item When the `other' regime is predicted, an asset should be sold (if it was bought) or bought back (if it was shorted).
\end{itemize}

\subsection{Performance metrics}\label{sec:performance_metrics}

\subsubsection{Financial performance metrics} \label{sec:financial_performance_metrics}

Three such metrics are used here, the \textit{annualised} (over a trading year of 252 days) \textit{Sortino ratio}, \textit{adjusted Sharpe ratio}, and \textit{information ratio}. The presented formulae will in each case make reference to the risk-free return $r_f$, but it should be noted that this quantity will in practice be conventionally set to zero. Furthermore, we note that $\mu(.)$ will denote the mean value of any daily-varying quantity, $\sigma(.)$ its variance, and $\sigma_{neg}(.)$ its negative semi-variance (downside deviation), with each of these calculated over all $t=1 \ldots T$ trading days in the test period. Each of the three metrics to follow is based on the calculation of a \textit{daily cost-adjusted trading return}, $r_M(t)$,

\begin{equation}\label{eq:cost_adjusted_returns}
r_M(t) = \sum _{\begin{array}{c} \mbox{assets} \\ i \end{array}}^{} w_{L,i}(t)r_{L,i}(t) + w_{S,i}(t)r_{S,i}(t),
\end{equation}

\noindent in which $r_{L,i}(t)$ is the return associated with longing (similarly, shorting) asset $i$ on day $t$, and $w_{L,i}(t)$ is the weight associated with longing (and similarly, for shorting) asset $i$ on day $t$. The weights are given by

\begin{equation}\label{eq:weights}
w_{L,i}(t) = \left\{ \begin{array} {rl} \frac{1}{n_{L,t}} & \mbox{  if asset } i \mbox{ is longed on day } t \\ 0 & \mbox{  otherwise} \end{array}\right. ,
\end{equation}

\noindent where $n_{L,t}$ is the number of assets longed (similarly, shorted) on day $t$. Note that an asset does not need to be traded, either longed or shorted, on a given day; such a situation is represented by setting $w_{L,i}(t) = w_{S,i}(t) = 0$. 

\begin{itemize}
    \item \textit{Annualised Sortino ratio} (ST). This is our primary metric, used for both hyperparameter tuning and out-of-sample performance assessment: 

    \begin{equation}\label{eq:sortino_ratio}
    ST = {\frac{\mu(r_M - r_f)}{\sigma_{neg}(r_M - r_f)}} \times \sqrt{252}.
    \end{equation}

    \item \textit{Annualised adjusted Sharpe ratio} (ASR) \cite{israelsen2005refinement}. This is a modification of the Sharpe ratio that treats positive returns in the same manner as the standard Sharpe ratio, but more strongly penalises negative returns:

    \begin{equation}\label{eq:ASR}
    ASR = {\frac{\mu(r_M - r_f)}{{\sigma(r_M - r_f)}^{\mu(r_M - r_f) / \mu|r_M - r_f|}}} \times \sqrt{252}.
    \end{equation}

    \item \textit{Annualised information ratio} (IR). This is the degree to which an actively managed portfolio, typically incurring significant transaction costs, can outperform a buy-and-hold investment in a benchmark index,

    \begin{equation}\label{eq:information_ratio}
    IR = {\frac{\mu(r_M - r_B)}{\sigma(r_M - r_B)}} \times \sqrt{252},
    \end{equation}

    \noindent where $r_B$ are benchmark returns. A value in excess of 1.0 is desirable (the managed portfolio then being more profitable than the index); over extended periods, such as the test period used here, this can be hard to achieve.
    
\end{itemize}

\subsubsection{Classification performance metric} 

Additionally to these financial metrics, the \textit{Matthews correlation coefficient}~\cite{matthews1975comparison} (MCC) has been used to check the robustness of the KMRF model as a classifier. The MCC was chosen for this purpose due to arguments that it is the most informative single-number metric that can be derived from the confusion matrix~\cite{chicco2020advantages}, especially for imbalanced data. We calculate the MCC separately for each class, as below, 

\begin{equation}\label{eq:mcc}
MCC_i = \frac{(TP_i \times TN_i - FP_i \times FN_i)}{\sqrt{(TP_i + FP_i) (TP_i + FN_i) (TN_i + FP_i) (TN_i + FN_i)}},
\end{equation}

\noindent in which $TP_i$, $TN_i$, $FN_i$, and $FP_i$ are true positives, true negatives, false negatives (incorrect assignments of a class $i$ example to any $j\neq i$), and false positives (incorrect assignments to class $i$ of any $j\neq i$ example), respectively, and $i \in \{bullish, bearish, other\}$.  The calculation of MCCs for each class is useful since the main objective is to predict bullish and bearish regimes and take trading actions based on these predictions; in contrast, the `other' regime is used only to close an open position, so the value of $MCC_{other}$ is of lesser importance.

\section{Results}

This section will examine the performance of the proposed KMRF model for each of the asset classes, during the relevant test periods 
(approximately four years in each case). It will first consider, in Section~\ref{sec:classification_performance}, the performance of the model purely as a classifier, making three-way regime predictions of bullish, bearish, or `other', with performance assessed by the MCC. It will then, in Section~\ref{sec:financial_performance}, move on to performance in relation to financially-motivated metrics, before finally, in Section~\ref{sec:feature_importances}, examining which of the many initially considered features listed in Section~\ref{sec:candidate_features} were discovered to be the most valuable for regime prediction. 
It should be noted that the KMRF results presented here are for the models in which BorutaShap feature selection has already been performed (on the basis of training data only) and the RF model then re-trained and re-optimised (again, on the basis of training data only, using the PGTS cross-validation method referenced in Section~\ref{sec:hyperparameter_optimisation}). This feature selection step reduced the numbers of features to 131 in the case of equities, 63 in the case of commodities, and 24 in the case of FX, the larger number of features used by the equities model being due to the larger number of candidate fundamental features in this case.

\subsection{Classification performance of the proposed KMRF model} \label{sec:classification_performance}

This section looks at the performance of the model as a classifier, with the Matthews correlation coefficient (MCC) (Eq.~\ref{eq:mcc}) used as a metric, for reasons explained in Section~\ref{sec:performance_metrics}.
It should be noted that a high MCC does not guarantee a high Sortino ratio, as the MCC is a counting measure that registers only whether a predicted regime label is correct or incorrect; a correct prediction may not be profitable if the market move is insufficient to outweigh the costs of trading. However, conversely, a high Sortino ratio might be achieved during the test period essentially by luck, but an MCC close to zero would expose this. Thus the dual use of MCC and ST is the best guide to the likely reliability of the model outside of its test period, going forward into the future.

Table~\ref{tab:mcc_scores} presents the result of this investigation. It is immediately apparent from the table that the MCCs for both the bullish and bearish regime labels are very high, in the context of financial prediction (for comparison, one can note that a prediction of closing price direction based on previous closing prices would typically not achieve an MCC greater than 0.05), while the MCCs for the `other' regime label are substantially lower. However, this latter is not a large problem as the `other' signal is used only to close a position, and hence a misclassification of an actually bullish or bearish regime to `other' would represent only a loss of profit due to the premature closing of a position, not a potentially much larger loss due to having taken the wrong position.

\begin{table}[] 
\caption {Test period MCC scores for the KMRF model for each asset class, presented separately for each of the three predicted regime labels; the average MCC over the bullish and bearish class labels is also given.}
\centering
\begin{tabular}{|l|c|c|c|c|}
\hline
\multicolumn{1}{|c|}{\textbf{Asset class}} & \textbf{\begin{tabular}[c]{@{}c@{}}Bullish\end{tabular}} & \textbf{\begin{tabular}[c]{@{}c@{}}Bearish\end{tabular}} & \textbf{\begin{tabular}[c]{@{}c@{}}Av. Bullish\\ \& Bearish\end{tabular}}& \textbf{\begin{tabular}[c]{@{}c@{}}`Other'\end{tabular}} \\ \hline
Equities                                   & 0.4809                                                                            & 0.3768                                                                   & 0.4289                                                                  & 0.1455                                                                 \\ \hline
Commodities                                & 0.4260                                                                            & 0.4191                                                                   & 0.4256                                                                     & 0.0632                                                                \\ \hline
Currencies (FX)                                & 0.5413                                                                              & 0.3295                                                                   & 0.4354                                                                     & 0.0552                                                                \\ \hline
\end{tabular}
\label{tab:mcc_scores}
\end{table}

\subsection{Trading performance: comparison of the proposed KMRF model with competitor models and benchmarks} \label{sec:financial_performance}

This section considers the result of using the generated regime predictions to trade. 
The KMRF model was compared to a hidden Markov model (HMM) and the KAMA+MSR model of~\cite{pomorski2023improving}. 
The HMM was picked as a comparison model due to its popularity for regime prediction, as evidenced in Section~\ref{sec:related_work}. 
The comparison with the KAMA+MSR model aimed to clarify the added value of predicting, as opposed to only detecting, as in~\cite{pomorski2023improving}, financial regimes.
The comparison was done in terms of financially-relevant metrics: the Sortino ratio (Eq.~\ref{eq:sortino_ratio}), used, as noted in Section~\ref{sec:hyperparameter_optimisation}, for hyperparameter optimisation; but also the adjusted Sharpe ratio (Eq.~\ref{eq:ASR}) and information ratio (Eq.~\ref{eq:information_ratio}), neither of which were used in any form of optimisation. 

Table~\ref{tab:metrics_comparison} shows that the KMRF model performed substantially better than its competitors during the four-year test periods, with respect to all these metrics. Notably, the KMRF model was the only one to achieve an IR~$> 1.0$, i.e., to be able to outperform the relevant index benchmark, for every asset class.
It should also be noted that predictions from all the models were both statistically significantly different from random and statistically significantly different from each other (at a significance level of 0.01 or better, in all cases).

\begin{table}[]
\caption {Test period financial metrics (annualised Sortino and adjusted Sharpe ratios, and information ratio (IR)) for the KMRF model in comparison to its competitor models (HMM, KAMA+MSR), for each asset class.}
\centering
\begin{tabular}{|l|l|l|l|l|l|}
\hline
\textbf{Asset class}                       & \textbf{\begin{tabular}[l]{@{}l@{}}Index\\benchmark\end{tabular}}                            & \textbf{Model}                        & \hspace{0.05 cm}\textbf{\begin{tabular}[l]{@{}l@{}}Sortino\\ ratio (ann.)\end{tabular}} & \hspace{0.05 cm}\textbf{\begin{tabular}[l]{@{}l@{}}Adj. Sharpe\\ ratio (ann.)\end{tabular}} & \hspace{0.05cm}\textbf{IR}\\ \hline

\multirow{3}{*}{Equities}         & \multirow{3}{*}{ACWI}                 & KMRF   & \hspace{0.05 cm}\textbf{24.99}\hspace{0.05 cm}                & \hspace{0.05 cm}\textbf{1.44}                        & \hspace{0.05 cm}\textbf{1.46}                                         \\ \cline{3-6} 
                                  &                                            & HMM           & \hspace{0.05 cm}0.32                 & \hspace{0.05 cm}0.005                        & \hspace{0.05 cm}-0.11 \hspace{0.02 cm}                                      \\ \cline{3-6} 
                                  &                                            & KAMA+MSR\hspace{0.05 cm}                      & \hspace{0.05 cm}-0.29                & \hspace{0.05 cm}-0.0004                      & \hspace{0.05 cm}-0.51                                        \\ \hline
\multirow{3}{*}{Commodities}      & \multirow{3}{*}{BBG\_Commodity} & KMRF & \hspace{0.05 cm}\textbf{12.66}                & \hspace{0.05 cm}\textbf{1.22}                         & \hspace{0.05 cm}\textbf{1.93}                                         \\ \cline{3-6} 
                                  &                                            & HMM           & \hspace{0.05 cm}-0.47                & \hspace{0.05 cm}-0.002                       & \hspace{0.05 cm}-1.27                                        \\ \cline{3-6} 
                                  &                                            & KAMA+MSR                      & \hspace{0.05 cm}0.24                 & \hspace{0.05 cm}0.002                        & \hspace{0.05 cm}-0.42                                        \\ \hline
\multirow{3}{*}{\begin{tabular}[l]{@{}l@{}}Foreign\\Exchange\end{tabular}} & \multirow{3}{*}{BBDXY}               &  KMRF                             & \hspace{0.05 cm}\textbf{4.72}                 & \hspace{0.05 cm}\textbf{0.29}                         & \hspace{0.05 cm}\textbf{1.45}                                         \\ \cline{3-6} 
                                  &                                            & HMM           & \hspace{0.05 cm}-0.37                & \hspace{0.05 cm}-0.0001                      & \hspace{0.05 cm}-0.88                                        \\ \cline{3-6} 
                                  &                                            & KAMA+MSR                      & \hspace{0.05 cm}0.17                 & \hspace{0.05 cm}0.1                          & \hspace{0.05 cm}-0.28                                        \\ \hline
\end{tabular}
\label{tab:metrics_comparison}
\end{table}

\begin{figure}
\centering
\includegraphics[scale=0.75]{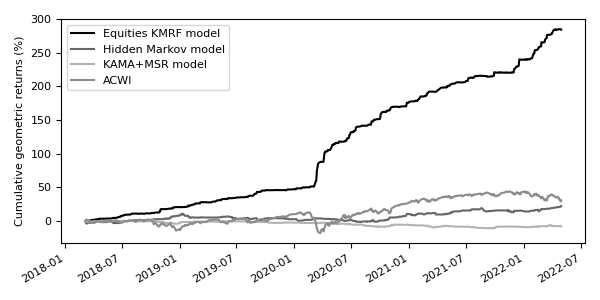}
\caption{Test period cumulative performance of the equities KMRF, HMM, and KAMA+MSR models, compared to a long-only position in the index benchmark.} \label{fig:equities_performance_chart_vs_benchmarks}
\end{figure}

\begin{figure}
\centering
\includegraphics[scale=0.75]{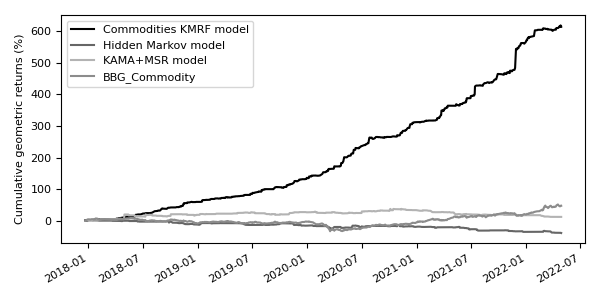}
\caption{Test period cumulative performance of the commodities KMRF, HMM, and KAMA+MSR models, compared to a long-only position in the index benchmark.} \label{fig:commodities_performance_chart_vs_benchmarks}
\end{figure}

\begin{figure}
\centering
\includegraphics[scale=0.75]{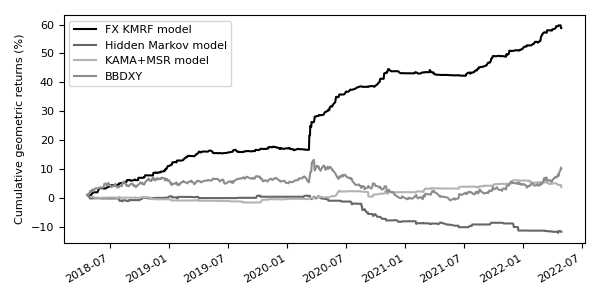}
\caption{Test period cumulative performance of the FX KMRF, HMM, and KAMA+MSR models, compared to a long-only position in the index benchmark.} \label{fig:FX_cumulative_return}
\end{figure}

The strength of the KMRF model is further illustrated in Figures \ref{fig:equities_performance_chart_vs_benchmarks}, \ref{fig:commodities_performance_chart_vs_benchmarks}, and \ref{fig:FX_cumulative_return}, for each of the asset classes, in terms of cumulative performance, 

\begin{equation}
Q(t) = \Bigr[ \prod _{s=1}^{t} (1+r_M(s)) \Bigl] \times 100,
\end{equation}

\noindent the wealth attained on day $t = 1 \ldots T$ by an investor with an initial ($t=0$) investment of \$100. This is compared with the performance of the competitor models (HMM and KAMA+MSR), as well as those of the appropriate index benchmarks. Note that the returns are cost-adjusted, apart from the index  benchmarks, and that the index benchmarks are not compounded, as they are simply bought at the start of the test period and sold at the end. The best-performing KMRF model was for the commodities asset class, but the outperformance of the KMRF model in the case of equities and FX is also very noticeable.

It may be recalled from Section~\ref{sec:contrarian_interpretation} that the KMRF model adopts a contrarian interpretation of the generated signals.
The HMM and KAMA+MSR models, in contrast, here use conventional interpretations of the trading signals.
It should be noted, however, that the competitor models were, as a further test, also run in contrarian mode; this resulted, for both competitors, for all three asset classes, in an even worse performance. Thus we feel confident it is the \textit{combination} of KMRF and contrarian trading that leads to the observed superior results.

\subsection{Feature importances and interpretation}\label{sec:feature_importances}

\begin{figure}
\includegraphics[width=\textwidth]{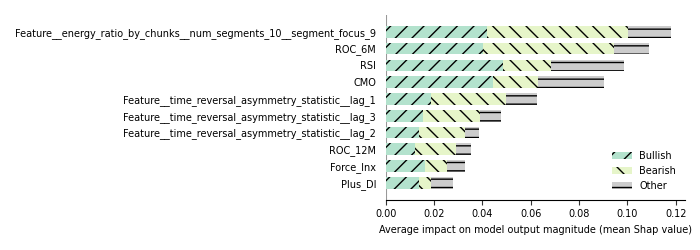}
\caption{Feature importances for the equities KMRF model.} \label{fig:equities_feature_importance}
\end{figure}

\begin{figure}
\includegraphics[width=\textwidth]{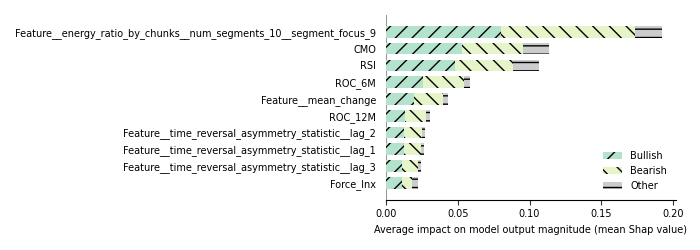}
\caption{Feature importances for the commodities KMRF model.} \label{fig:commodities_feature_importance}
\end{figure}

\begin{figure}
\includegraphics[width=\textwidth]{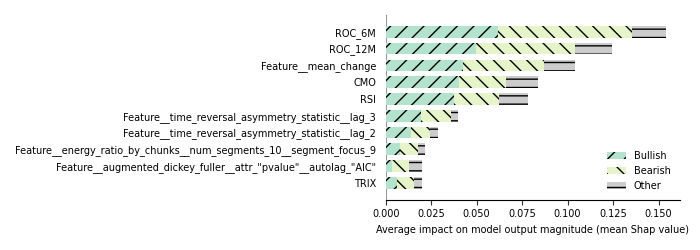}
\caption{Feature importances for the FX KMRF model.} \label{fig:FX_feature_importance}
\end{figure}

Figures \ref{fig:equities_feature_importance}, \ref{fig:commodities_feature_importance}, and \ref{fig:FX_feature_importance} show the top 10 most important features out of those selected by the BorutaShap algorithm, in terms of their absolute impact on the predictions of the equities, commodities, and FX, respectively, KMRF models, with the impact of each feature on prediction being computed using Shapley values~\cite{shapley1953value}. The figures present feature importances only for the 15\% test sets, as this is more relevant in relation to out-of-sample use than the same statistics for the training set. It can be noted that all the models for all three asset classes are driven by a similar set of features, though the FX model, unlike the others, is not primarily driven by the `energy ratio by chunks' feature. The largest impact on all models’ output probabilities is from technical momentum variables. Macroeconomic variables have a lesser effect, which is in line with the data frequency (daily); in the financial industry, it is commonly known that technical features are more predictive for higher frequency data, while macroeconomic features conversely tend to be more predictive for lower frequency data, such as monthly. This is because macroeconomic variables tend to change more slowly than technical indicators, as well as usually lagging current market conditions.

\section{Discussion and conclusions}

The work of this paper has built upon a previous work~\cite{pomorski2023improving} which proposed a novel method of detection of regimes, referred to as the KAMA+MSR model, for financial time series. This model combined the Markov-switching regression (MSR) model's ability to detect changes in volatility with the ability of Kauffman's adaptive moving average (KAMA) to detect an upward or downward trend.
However, the KAMA+MSR model suffered from inevitable lag-related losses associated with regime detection, as opposed to prediction,
and thus this work has focused on predicting the KAMA+MSR regime classes ex ante, using a random forest classifier, resulting in a regime prediction model we term the \textit{KMRF} (KAMA+MSR+RF) model.
Three major asset classes were considered: equities, commodities, and foreign exchange (FX), to increase the robustness of our conclusions. Results over a four-year test period evidenced the success of the proposed methodology, as each of the KMRF models, based on its generated signals, was able to outperform the selected benchmarks, as well as generate substantial risk-adjusted returns, net of costs, with the commodities KMRF model having the highest cumulative geometric returns, while the equities KMRF model achieved the highest annualised Sortino and adjusted Sharpe ratios.
However, one major finding of this work was that the KMRF models provide signals best interpreted as contrarian:
it was discovered the higher the probability of a bullish regime, the more likely it was that the asset price would actually fall, while the higher the probability of the bearish class, the more likely it was that the asset price would in fact rise. 
Adopting this interpretation gave rise to substantial and sustained profits for all three asset classes, during our four-year test period, though
the profits thereby obtained do come at a certain cost, as a large number of profitable trades are due to short positions. Even though frequent shorting is not impossible, it could be problematical, as shorting may be constrained by institutions providing brokerage services on grounds of financial risk, or be excluded from consideration by investors for ethical reasons.

Turning to future work, the KMRF model could be enhanced with dynamic asset weight allocation (instead of using mean weights, as in Eq.~\ref{eq:weights}), as well as a more complex cost model (this work having assumed the fixed costs of Table~\ref{tab:transaction_costs}, independently of the liquidity and volatility of the traded assets). 
To overcome the above-discussed issues related to shorting, one possibility would be to buy put options; however, such options incur additional costs, as premia must be paid for the right to sell short the underlying asset. A more attractive and widely-adoptable possibility might be to focus on a long-only portfolio strategy, one which would also be capable of dynamically shifting asset weights, depending on the underlying asset return estimates, in order to construct high-Sortino long-only portfolios under a realistic cost model, and such work is ongoing.


\bibliographystyle{unsrt}  
\bibliography{references}

\end{document}